\documentclass[%
 aip,
 rsi,
floatfix,%
 amsmath,amssymb,
 reprint,%
]{revtex4-1}

\usepackage[utf8]{inputenc}
\usepackage[T1]{fontenc}

\usepackage{graphicx}
\usepackage{dcolumn}
\usepackage{bm}

\begin{document}

\preprint{AIP/123-QED}

\title[Power regulation and electromigration]{Power regulation and
electromigration in platinum microwires}

\author{Ottó Elíasson}
\affiliation{Science Institute, University of Iceland, Dunhagi 3, Reykjavík IS-107, Iceland}
\author{Gabriel Vasile}%
\affiliation{Science Institute, University of Iceland, Dunhagi 3, Reykjavík IS-107, Iceland}
\affiliation{National Institute of Research-Development for Cryogenics and Isotopic Technologies, Uzinei 4, Ramnicu Valcea RO-1000, Romania}
\author{Sigurður Ægir Jónsson}
\altaffiliation{Present address: Computational Biology and Biological Physics, Lund University, Sölvegatan 14A, SE-223 62 Lund, Sweden}
\affiliation{Science Institute, University of Iceland, Dunhagi 3, Reykjavík IS-107, Iceland}
\author{G.~I.~Gudjonsson}
\affiliation{Science Institute, University of Iceland, Dunhagi 3, Reykjavík IS-107, Iceland}
\author{Mustafa Arikan}%
\affiliation{Science Institute, University of Iceland, Dunhagi 3, Reykjavík IS-107, Iceland}
\author{Snorri Ingvarsson}%
 \email{sthi@hi.is.}
\affiliation{Science Institute, University of Iceland, Dunhagi 3, Reykjavík IS-107, Iceland}

\date{\today}

\begin{abstract}
We introduce a new experimental setup with a biasing circuit and computer
control for electrical power regulation under reversing polarity in Pt
microwires with dimensions of 1$\times$10 $\mu$m$^2$. The circuit is computer
controlled via a data acquisition board. It amplifies a control signal from the
computer and drives current of alternating polarity through the sample in
question. Time-to-failure investigations under DC and AC current stress are
performed. We confirm that AC current stress can improve the life time of
microwires at least by a factor of $10^3$ compared to the corresponding
time-to-failure under DC current stress.
\end{abstract}

\maketitle

\section{\label{sec:inng}Introduction}
Our recent research has involved studying radiative properties of electrically
heated microwires in the
infrared~\cite{Ingvarsson:2007vga,Au:2008hy,Renoux:2011kd,Vasile:2012}.  We
often refer to these as microheaters.  Biasing with a larger current raises the
temperature of the microheater and thus increases the intensity of radiation. It furthermore shifts the radiation spectrum to shorter wavelengths, much like for an ideal
blackbody, as described by Planck's law\cite{Born:2005}. The details of the
spectrum and the wire's resistance at given bias conditions depend on each
microheater's physical dimensions. Thus obviously it is important to regulate
the temperature in our microheaters.
\begin{figure}
	\includegraphics[width=\columnwidth]{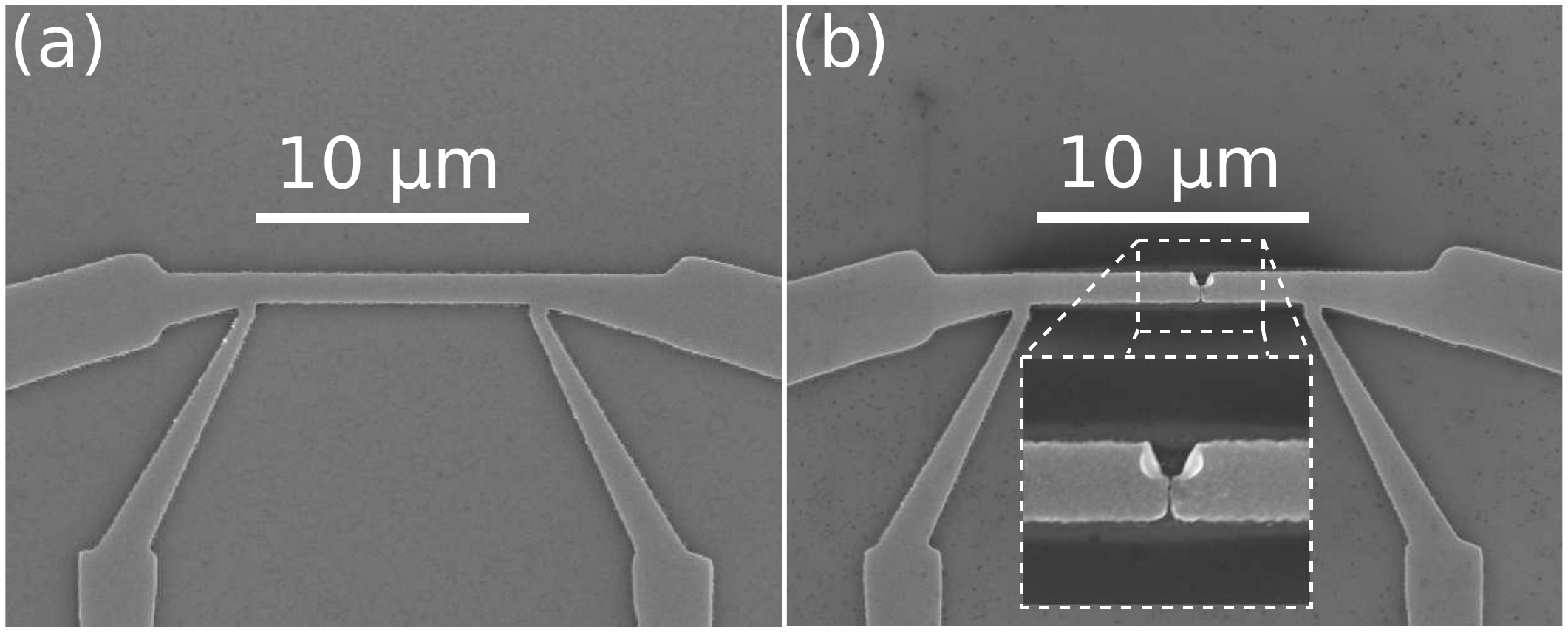}
	\caption{Scanning electron micrographs of microheaters used in our study
	(a) before, and (b) after onset of electromigration. A change in
	the substrate's color surrounding the heater due to heating is apparent.
	Inset: Englargement of the failure location showing the ``mousebite''
	void, and from it, a narrow slit across to the opposite edge of the
	wire.}
	\label{fig:heater_broke}
\end{figure}

We have found that
to a good approximation the temperature of our microheaters, $T_{\rm h}$,
depends linearly on the power $P_{\rm h}$ dissipated (Joule heating) in
them:~\cite{Ingvarsson:2007vga,Jonsson:2009wb}
\begin{align}
	T_{\rm h} = T_0 + \frac{dT}{dP} P_{\rm h}~\quad,\label{eq:temp_power}
\end{align}
where $T_0$ is room temperature (i.e. $T_{\rm h}$ at $P_{\rm h} = 0$ mW). 
The thermal impedance of the microheater, $dT/dP$, is governed by its
surroundings, i.e.\ the thermal properties of its environment,
and remains constant even upon thermal cycling. The resistance of the heaters
$R_{\rm h}$ varies with temperature. Within a narrow enough temperature range
around or above room temperature, it may be approximated with the well known relation
\begin{align}
	R_{\rm h} = R_0(1 + \alpha \Delta T), \label{eq:res_temp}
\end{align}
where $\alpha$ is the temperature coefficient of resistance, $\Delta T = T_{\rm
h} - T_0$ and $R_0$ is the resistance at room temperature.  The electrical
resistance is a property of the wire, and depends on its geometry and the
material's resistivity. Also we have found that $\alpha$ may in some cases
change upon thermal cycling, presumably due to alterations in grain size
or other annealing effects.  However, the significance of
eq.~(\ref{eq:temp_power}) is that it states that if you know the power
dissipated in your wire, you can tell its temperature.  One can measure $dT/dP$
by current-voltage measurements in a narrow range where $\alpha$ remains
unchanged, and subsequently expose the heater to a much broader range of power
and temperature confident that $dT/dP$ remains constant even though $\alpha$ may
change.

Increased current results in greater Joule dissipation and higher temperature.
However, there are limits, and as one reaches a critical current density
($J\sim10^7$ A/cm$^2$) void formation and electromigration will cause the wire to break, resulting
in an open electrial circuit~\cite{Lienig:2005jj}. Electromigration is the process of current-induced
self-diffusion in metal lines, and it is the mechanism responsible for long-term
wear-out of interconnects in integrated circuits
\cite{Black:1969fc,Gardner:1987ky,Clement:2001bu,Banerjee:2001cn,Lloyd:1997tk,Lloyd:1999tz}.
Figure~\ref{fig:heater_broke}
displays a typical failure in one of our wires; i.e.\ there is a ``mousebite''
void formation and a narrow slit extending from it across the width of the
wire\cite{Orio:2010,Oates:1993}.
With the carefully tapered contact design, failures typically occur at or near
the middle of the length of the wire, with void formation at the edge.  The
highest temperature occurs in the middle of the wire, but the largest temperature
gradient, in-plane perpendicular to the wire, is at the edge midway along the
length\cite{Jonsson:2009wb}.  This causes large mechanical stress at the edge.
For our purposes it is very useful to extend the time-to-failure at a given
current bias of our microheaters as much as possible. The time-to-failure of the
wire $\tau$, can be greatly increased by reversing the current
polarity at a frequency $f$ much greater than that corresponding to the
time-to-failure of the heater under DC current stress $\tau_{\rm DC}$
\cite{Liew:1989ds,Tao:1993ck,Tao:1993cm}. 
Our need to maintain constant temperature, thus constant power dissipation,
while extending the time-to-failure of our heaters is somewhat unique. It is
complicated by the fact that there are slow variations in the wires' resistance
values (as discussed below). We were unable to find commercial solutions for
this. While communications via the IEEE standard GPIB are sufficiently fast to
deal with slow changes in resistance, they are too slow to allow rapid switching
of polarity. D/A converters provide speed but do not output adequate power in
all cases.
Thus a special biasing circuit was designed and
built, that enables measurement of the power dissipation in the heater itself as
well as allowing rapid change of bias polarity.  We used computer control to
regulate the power and to reverse the current polarity. By reversing
the current polarity we also expected to be able to nondestructively reach
higher heater temperatures $T_{\rm h}$ than before.

The outline of this paper will be as follows. We introduce the experimental
setup including the circuit and the computer control (the program used for power
regulation under AC stressing). We present results of measurements on the
time-to-failure of the microheaters comparing AC and DC current stress and
discuss some distinct features observed in the heaters' resistance, $R_{\rm h}$,
as a function of time.

\section{Experimental Setup}
To bias our microheaters we use a probe station and microprobes to connect to
the sample. We employ our new biasing circuit and a programmable National
Instruments Data Acquisition board (DAQ-board) of type NI-USB 6229 to regulate
the power dissipation in the heater (and thus its temperature) via computer
control. The biasing circuit is powered by $\pm V_{\rm CC}=\pm15$~V supply
voltage. 

\subsection{Biasing circuit}
\begin{figure}[ht]
	\includegraphics[width=\columnwidth]{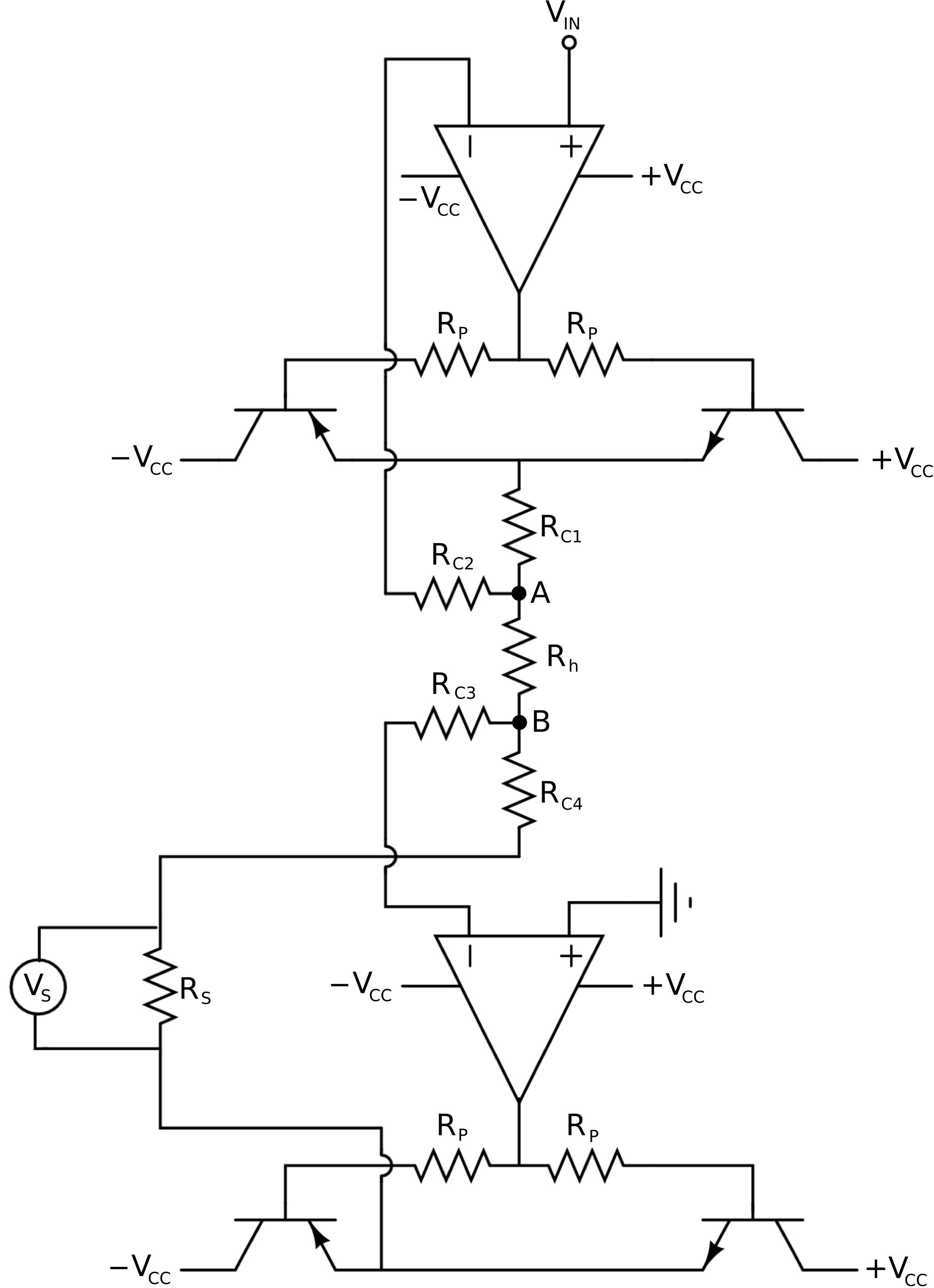}
	\caption{\label{fig:prc} The circuit that enables monitoring the power
	dissipation in a microheater and allows rapid switching of bias polarity.}
\end{figure}

The circuit in fig~\ref{fig:prc} (initially described in ref.~\onlinecite{Jonsson:2009wb}) is
essentially composed of two followers, each comprised of an operational
amplifier (TL072 in our case) and a push-pull transistor amplifier (C557B pnp
and C547B npn transistors).  The resistors $R_{\rm P}$ at each tranistor's base
are to limit the voltage seen by the base.
We use a four point probe connection to our heaters, i.e.\ two providing current
and two to sense voltage. Their contact resistances are represented by $R_{\rm
C1}$, $R_{\rm C2}$, $R_{\rm C3}$ and $R_{\rm C4}$ respectively.
The microheater is represented by the resistance $R_{\rm h}$.
Since the inputs of op-amps draw almost no current, the voltage drop in our
voltage sensing leads/contacts ($R_{\rm C2}$ and $R_{\rm C3}$) is negligible and the current $I_{\rm
h}$ runs directly through contacts $R_{\rm C1}$ and $R_{\rm C4}$. The followers
ensure that the voltages at points $A$ and $B$ equal the input voltages $V_{\rm
in}$ and 0, respectively. Thus we have the relation $I_{\rm h}=V_{\rm
in}/R_{\rm h}$. However, as the microheater's resistance increases as it is
heated, the current is measured via the voltage drop across a known precision
resistor $R_{\rm S}$. The power dissipation may then be found by 
\begin{align}
	P_{\rm h} = I_{\rm h}V_{\rm h} = \frac{V_{\rm S}}{R_{\rm S}}V_{\rm in}, \label{eq:pow_heat}
\end{align}
all known quantities.

It is important to note that if a microheater sample is removed and reconnected
there is a high risk of transitents, potentially destructive to the sample.  For
this reason we turn off the supply voltage and ramp it up in a controlled manner
once the microheater is in place.

\subsection{Computer control}
The DAQ-board, which manages the signals to ($V_{\rm in}$) and from ($V_{\rm
S}$) the biasing circuit, was programmed in Microsoft's Visual C++ environment
for fast response.  As the current through the microheater is ramped up we start
with a regular proportional ``P''-control, where the desired output is manipulated by
adjusting the voltage $V_{\rm in}$ proportional to the ``error signal'' $P_{\rm
h}-P_{\rm d}$, where $P_{\rm d}$ is the desired power value.  When $P_{\rm h}$
approaches the desired value $P_{\rm d}$ a simpler “fuzzy logic” control takes
over and maintains $P_{\rm h}$ at the desired value. An advantage of the fuzzy
logic control is that it is not computationally intensive, thus fast.  This can
easily be replaced by a control of the user's choice, PID or more advanced
methods. 

In order to combat
electromigration we reverse the polarity of the voltage $V_{\rm in}$ at a
frequency $f = 20$~kHz, forming a square wave.  This frequency was chosen based
on ref.~\onlinecite{Tao:1993ck} (where the current is regulated, but not the
power as in our case), that showed that there was a three order of magnitude
lifetime gain by reversing polarity in Al wires at this rate.  However,
increasing the frequency by another three orders of magnitude gave a very modest
gain in lifetime. Our power regulation, on the other hand, is done at a much
lower frequency $f_r = 200$ Hz. This regulation amounts to adjusting the
amplitude of the square biasing wave in response to the changes in the
microheater's resistance.  This much lower frequency of regulation is quite
adequate, since the changes in microheater resistance are mostly slow compared
with $f_r^{-1}$, as can be seen in fig.~\ref{fig:res_time}.  Although this
frequency may be raised somewhat, it is limited by communication and computation
time.  With this simple method we can easily maintain the power dissipation
within 0.1\% of the desired value.

\section{Sample properties}
\begin{figure}[t]
	\includegraphics[width=\columnwidth]{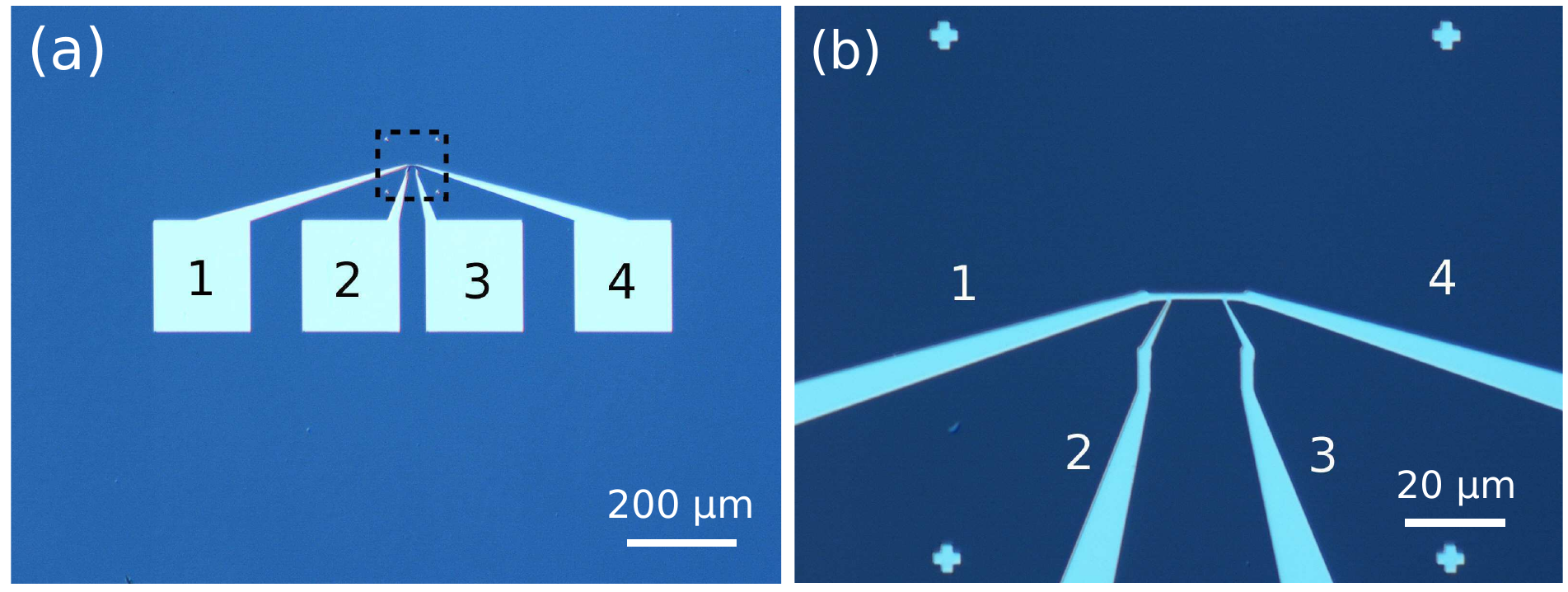}
	\caption{Optical microscope images of heaters used in this study. The
	contact pads are labeled with numbers 1--4. Each pad is
	180$\times$200~$\mu$m$^2$.  The dotted line in fig.~(a) demarcates the
	area pictured in fig.~(b).}
	\label{fig:heater}
\end{figure}

The microheaters used in our study were manufactured on a Si/SiO$_2$ substrate
with e-beam lithography to ensure good shape definition.
The heaters, typically about 60 on each
substrate chip, are deposited via DC magnetron sputtering. Their structure is
simple, a 5 nm polycrystalline Cr
adhesion layer followed by a 50 nm polycrystalline Pt layer. The details of our
samples are described in ref.~\onlinecite{Au:2008hy}.  In the present study all the
microheaters are of the same lateral dimensions 1$\times$10 $\mu$m$^2$.
Samples are depicted in fig.~\ref{fig:heater}.
In order to make sure that the heaters were as comparable as possible, they were
characterized by standard electrical measurements before our experiments. Three
quantities describing the properties of our heaters are of special interest.
These are the resistance $R_{\rm h}$ (in particular the cold resistance $R_c$),
the thermal impedance $dT/dP$ and the temperature coefficient of resistance
$\alpha$ (prior to thermal cycling, used in the process of determining
$dT/dP$).

An {\em as-grown} (untouched) microheater had a cold resistivity of  
\begin{align}
	\rho_c = 20 \ \mu\Omega\,\text{cm}, \label{eq:resist}
\end{align}
which is about 90\% greater than many reports of bulk resistivity values of Pt
in the literature\cite{kittel}, $\rho_{\rm bulk}$~=~10.4~$\mu\Omega\,$cm. We
attribute this deviation primarily to surface scattering, and grain boundary
scattering.  The TCR value, $\alpha$ was obtained by measuring resistance of
heaters sitting on a hot plate whose temperature is
increased stepwise resulting in corresponding changes in resistance. We measured over a
temperature range from room temperature up to about 110$^\circ$C, and found
$\alpha$ from fitting eq.~(\ref{eq:res_temp}), to be
\begin{align}
	\alpha = 2.20\times10^{-3} \text{ K}^{-1}.
\end{align}
Plotting resistance versus power allows an estimation of $dT/dP$. For our
heaters we obtained
\begin{align}
	\frac{dT}{dP} = 4.7 \text{ K/mW}. \label{eq:dTdP}
\end{align}

After such characterization one heater was subjected to a constant high power
for a few minutes, long enough to observe an obvious decrease in the
heater's resistance, see below.  Following
this procedure the heater's $I$-$V$ profile is obtained again to measure the
cold resistance, and its $\alpha$ value is measured again to see if it had
changed. No changes were found in either $\alpha$ or $dT/dP$ in our
measurements. 

\section{Time-to-failure measurements}
\begin{figure*}[ht]
	\includegraphics[width=\textwidth]{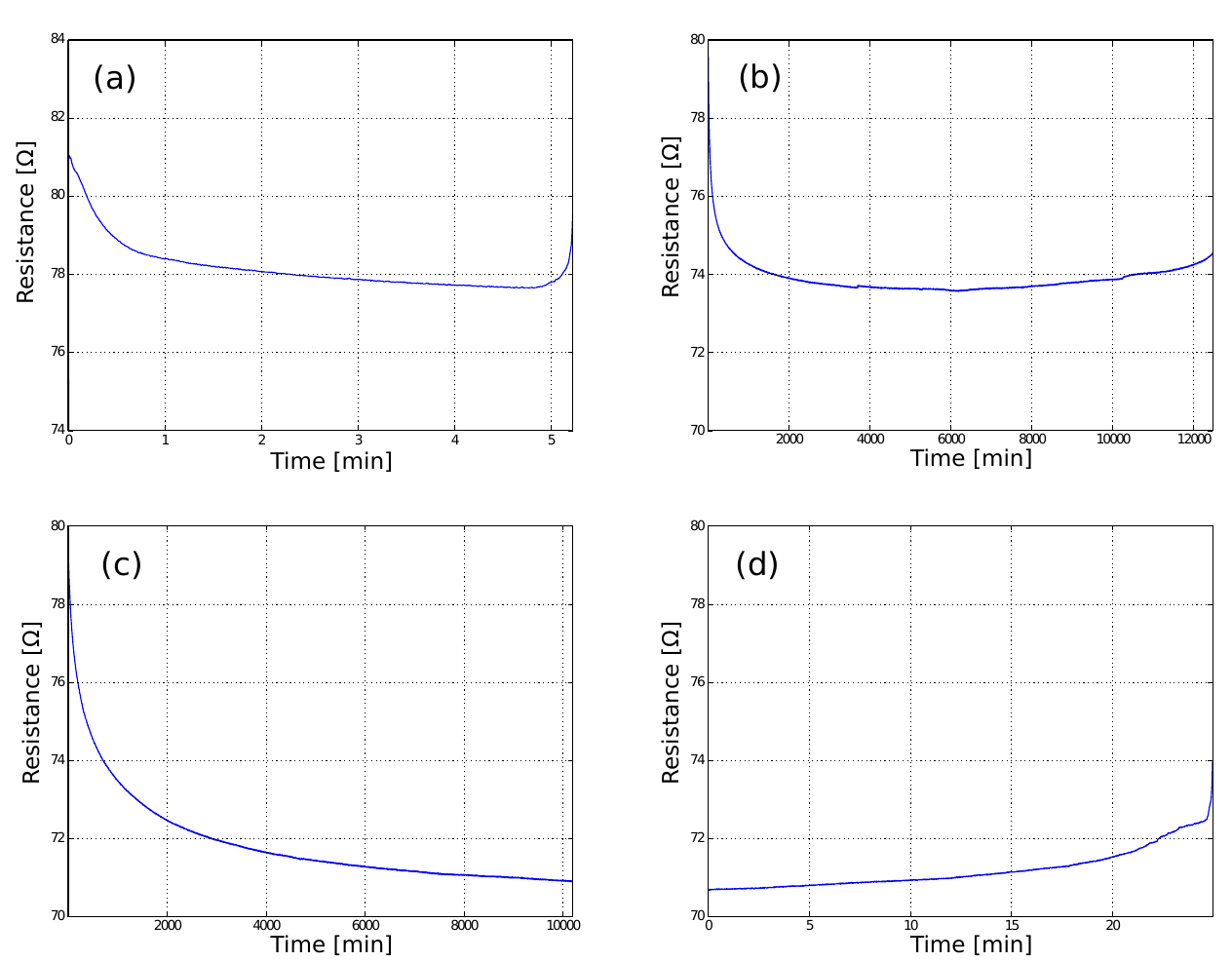}
	\caption{$R$-$t$ graphs of representative microheaters used in the
	research. (a) $P_{\rm h} = 100$ mW under DC current stress. (b) $P_{\rm
	h} = 100$ mW under AC current stress. (c) and (d) AC and DC
	measurements, respectively, of the same microheater at $P_{\rm h} = 90$
	mW.  The AC measurement was terminated and continued under DC bias.}
	\label{fig:res_time}
\end{figure*}
At low current density and low power the lifetime of our microheaters is, for
practical purposes, infinite.  However, as bias is increased their lifespan
shortens rapidly.  We wished to compare time-to-failure of heaters at fixed
power under DC and AC stress to see if bias polarity reversal would help.  We
regulated the power both at a power of $P_{\rm h}=90$~mW and at $P_{\rm
h}=100$~mW, corresponding to about 440$^\circ$C and
490$^\circ$C respectively according to eq. (\ref{eq:temp_power}) and
(\ref{eq:dTdP}). Typical results are displayed in fig.~\ref{fig:res_time}~(a)
showing resistance as function of time under DC stress at $P=100$~mW, and (b)
another identical microheater at the same power but under AC stress, i.e.\ a
square wave with $f=20$~kHz. Both show an initial gradual and smooth drop in
resistance over a signifant portion of the lifetime, that may be viewed as a
``repair process'' for the wire.  After reaching a minimum the resistance starts
to increase, with occasional steps, presumably due to movement of material,
i.e.\ electromigration.  Finally, there is a sharp rise in resistance and the
wire breaks.  Note that the time-to-failure in (b) is more than two orders of magnitude
longer than in (a).  Measurements of the DC time-to failure of 23 samples resulted in the following mean time-to-failure for 90~mW and 100~mW
respectively:
\begin{align}
	\tau_{\rm DC,90} &= [4000 \pm 200] \text{ s} \\
	\tau_{\rm DC,100} &= [270 \pm 30] \text{ s}~.
\end{align}

Measuring the time-to-failure of the heaters under AC current stress turned out
to be very time consuming as their lifetime was extended significantly. Two
heaters were measured under AC current stress at $P_{\rm h}~=~90$~mW and three
were measured at $P_{\rm h}~=~100$~mW. 
\begin{table}
\caption{\label{tab:table1} Results of time-to-failure measurements of microheaters under AC current stress. Numbers are assigned to measurements for text references.}
\begin{ruledtabular}
\begin{tabular}{lccc}
\ no. & $P_{\rm h}$ [mW] & $\tau_{AC}$ [hrs] & $\tau_{AC}/\tau_{DC}$\\
\hline
\ 1\footnotemark[1] & 90  & 170 & 153\\
\ 2\footnotemark[1] & 90  & 137 & 123\\
\ 3 & 100 & 208 & 2800\\
\ 4 & 100 & 122 & 1600\\
\ 5 & 100 & 129 & 1700\\
\end{tabular}
\end{ruledtabular}
\footnotetext[1]{These measurements were aborted prior to heater destruction.}
\end{table}
The results are listed in table \ref{tab:table1}. It is apparent from the values
in the last column of table \ref{tab:table1}, which holds comparison of the
time-to-failure of microheaters of AC and DC current stress, that the
time-to-failure of the heaters increased at least by a factor of $10^3$. The
heaters measured under 90~mW power were stopped manually since it would have
taken about 7 weeks of measuring each heater to wait for a thousandfold increase
of the time-to-failure, assuming a similar lifespan extention as for the 100~mW
case.  The results of our measurements display a dramatic increase in the
time-to-failure of the microheaters under AC current stress compared to DC
current stress, in agreement with ref.~\onlinecite{Tao:1993ck}.

Figure \ref{fig:res_time}~(c) displays results of a 90~mW measurement under AC
stress.  It displays the same initial behavior as described above, but it was
terminated after about 7 days, the resistance slope still gradually decreasing.
Based on the shape of the resistance curves and the location of the resistance
minima with respect to breaking time one could predict that under these
conditions this particular microheater would have lasted at least another week.
However, we decided to expose this same microheater to DC bias to see how it
fared.  This is shown in fig.~\ref{fig:res_time}~(d).  It lasted almost
25~minutes, i.e.\ about a third of the mean time-to-failure of as-grown samples at 90~mW and DC bias.  Note how the simple act of turning
off the AC bias and turning on again a DC bias abruptly changes the slope of
resistance versus time from negative to positive.  Also note that the initial
resistance in (c) is around 80 $\Omega$, while in (d) it starts at 71~$\Omega$,
i.e.\ where it left off in (c).
This shows that the changes in resistance are permanent, i.e.\ an
irreversible change takes place in the wires.  This contradicts previous
results, e.g.\ in ref.~\onlinecite{Lin:2013} on Cu interconnects.  We belive this is caused by grain-growth
enabled by self-heating annealing in our wires. This effect is studied further
in ref.~\onlinecite{Otto:2014}.

\section{Conclusions}
We have built a biasing circuit that is computer controlled to regulate power
dissipation in microwires, i.e.\ microheaters. This is important in order to maintain a constant
temperature of the heaters, to stabilize emission intensity and radiation spectrum.  
We have used our setup to investigate the electromigration affecting our
microheaters in terms of DC and AC time-to-failure measurements. Heaters were
regulated both using power $P_{\rm h}$ = 90 mW and $P_{\rm h}$ = 100 mW,
corresponding to about 440$^\circ$C and 490$^\circ$C, respectively and were 
measured both under DC and AC current stress. While the measurements at $P_{\rm
h}$ = 90 mW proved unfeasible due to long lifetime in the AC case, the measurements at $P_{\rm
h}$ = 100 mW showed that the AC time-to-failure of microheaters increased at
least by a factor of $10^3$ compared to the corresponding DC time-to-failure.

\begin{acknowledgments}
This work was partially supported by The Icelandic Research Fund grant
no.~120002023. We would like to thank Pauline Renoux for help with obtaining the
SEM images.
\end{acknowledgments}

%
\end{document}